\documentstyle[preprint,revtex,eqsecnum]{aps}

\tightenlines

\begin{document}

\draft


\begin{title}

Escape and spreading properties \\
of charge-exchange resonances in ${}^{208}$Bi

\end{title}

\author{G. Col\`o, N. Van Giai}
\begin{instit}  Division de Physique Th\'eorique, Institut de Physique
Nucl\'eaire,\\ 91406 Orsay Cedex, France \end{instit}

\author{P.F. Bortignon}
\begin{instit} Dipartimento di Fisica, Universit\`a degli Studi,\\ and
INFN, Sezione di Milano, Via Celoria 16, 20133
Milano, Italy \end{instit}

\author{R.A. Broglia}
\begin{instit} Dipartimento di Fisica, Universit\`a degli Studi,\\ and
INFN, Sezione di Milano, Via Celoria 16, 20133
Milano, Italy,\\ and Niels Bohr Institute, University of
Copenhagen, 2100 Copenhagen, Denmark \end{instit}

\begin{abstract}
The properties of charge-exchange excitations of ${}^
{208}$Pb with $\Delta L = 0$, i.e., the isobaric analog and Gamow-Teller
resonances, are studied within a self-consistent model making use of
an effective force of the Skyrme type.
The well-known isobaric analog case is used to assess the reliability
of the model. The calculated properties of the
Gamow-Teller resonance are
compared with recent experimental measurements with
the aim of better understanding the microscopic structure of this mode.
\end{abstract}

\pacs{IPNO/TH 94-22 \begin{flushleft} \hfill{APRIL 1994} \end{flushleft}}

\narrowtext

\section{Introduction}\label{sec:intro}

Giant resonances can
be described as coherent superpositions of particle-hole excitations
and their particle decays into given hole channels provide insight about the
corresponding particle-hole amplitudes. Thus, the study of decay properties
is a unique way to test model predictions for the resonance wave function.
Experimental studies have recently been made of the particle decay properties
of giant resonances by means of exclusive measurements where inelastically
scattered projectiles, or light products of transfer reactions,
are detected
in coincidence with nucleons emitted by giant resonances of the target system.
The corresponding results provide severe tests for the theoretical
models.

The Gamow-Teller resonance (GTR) is the manifestation of the spin-isospin
nuclear zero sound. This mode was predicted theoretically
as the mechanism for explaining
the missing strength in the decay of $\beta$-radioactive nuclei
\cite{Ike6365}. Its subsequent observation in (p,n) reactions has opened an
important chapter of the historical development of experimental study of
giant resonances when the availability of higher energy projectiles
in the early 80's led to the realization of reaction
conditions under which spin-isospin modes can
optimally be excited \cite{Gaa81a}.
A first experiment aimed to study the proton decay of the GTR was performed
in Groningen, and reported in Ref. \cite{Gaa81b}. The results of this
measurement raised a number of questions, as they seemed to imply a
vanishing spreading width for the GTR, which seems quite unlikely.

Renewed interest in the study of the properties of the GTR is indeed
testified by two recent measurements,
with different characteristics, at Osaka and MSU. The proton decay of
the resonance in ${}^{208}$Bi has been investigated, and in the Osaka
experiment \cite{Aki94}, the results
appear in modest agreement with previous theoretical estimates given in
Refs. \cite{Gia89} and
\cite{Mur_two}, whereas preliminary indications from the MSU
experiment seem to give a
somewhat different result \cite{Gal_pr}.
This
has motivated the present work, in which we apply
to the charge-exchange modes a model of giant resonances which includes
in a self-consistent way the coupling mechanisms leading to the
damping of these modes. This model has been already used
to study the properties of the isoscalar giant monopole
resonance in ${}^{208}$Pb \cite{Col92}.

We have first concentrated, as a test case, on the isobaric analog
resonance (IAR) in the nucleus ${}^{208}$Bi where the data provides
a clear picture of the resonance properties.
The IAR
would be degenerate with its isobaric multiplet partner, the ground state
of ${}^{208}$Pb, if the nuclear Hamiltonian were completely charge-
independent. The energy difference, due mainly to Coulomb effects, has been
studied over a long period of time and it is accurately known. The width
of the IAR is known to be very small. This is because it has
the same isospin as the parent state while surrounding states have
the isospin of the ground state of ${}^{208}$Bi, which differs by one unit.
Consequently, they couple only weakly with the IAR. Thus, excitation energy and
width of this resonance provide a stringent test for any theoretical model of
charge-exchange collective states.

We shall show how this test has been passed by our model in
Sec.\ \ref{sec:iar},
while the predictions associated with the GTR are presented in
Sec.\ \ref{sec:gtr}.
Conclusions are drawn in Sec.\ \ref{sec:conclu}.
Before this, we discuss in Sec.\ \ref{sec:gen} the general features
of the model already
applied to other collective modes \cite{Col92} while more specific
considerations relevant to charge-exchange modes are made in
Sec.\ \ref{sec:chex}. Detailed expressions for the Hamiltonian matrix and decay
branching ratios are given in Appendices A and B.

\section{General formalism}\label{sec:gen}

The Random Phase Approximation (RPA) provides, as a rule, an accurate
description of the centroid of giant resonances and the fraction
of the energy weighted sum rule (EWSR) exhausted by the mode.
However,
it is necessary to go beyond this approximation scheme in order
to explain the damping properties of the collective motion.
Indeed, giant
resonances are known experimentally to have an energy width and
therefore,
a finite lifetime. A general theory of the resonance width can be found,
e.g., in Refs. \cite{Brt83,Yos83}, and a list of references
of much of the theoretical work carried out during the last decade
can be found in \cite{Spe91}. In many of these calculations
parameters were
adjusted at various steps of the computational procedure in order to
fit experimental data. In what follows we present a model in which
a phenomenological
effective interaction of the Skyrme type is used as the starting point
but after this {\em no further tuning of the parameters} is introduced
up to final results.

We start by solving the Hartree-Fock (HF) set of equations for a given
nucleus $(A,Z)$ using a Skyrme two-body interaction
\cite{VauBei}. A set of occupied single-particle levels is then obtained,
as well as a self-consistent mean field, which is then diagonalized
on a basis made up with
harmonic oscillator wave functions. This diagonalization procedure can
also be replaced by solving the HF mean field with a box boundary
condition.
It provides a discrete set of occupied and unoccupied levels,
many of which are at positive energy. The quasi-bound levels belong to
this set. After choosing a
convenient cutoff - details about numerical procedures are given
in Sec.\ \ref{sec:chex} - a finite set of occupied and unoccupied levels
labeled by $|i\rangle$ is determined. In the following, we denote
the corresponding energies and wave functions by $\varepsilon_i$ and
$\varphi_i$ (the radial part of the wave function will be expressed as
$u_i(r)$).

Let us call $Q_1$ the subspace of nuclear configurations made up with
the HF ground state and all the possible one particle-one hole
(1p-1h) excitations built
within the set $|i\rangle$. We shall denote by
the same symbol subspaces and projectors onto them. Now, the nuclear
Hamiltonian can be written as
\begin{equation}
    H = H_0 + V_{ph},
\label{H_rpa}\end{equation}
where $H_0$ is the HF Hamiltonian and $V_{ph}$ is the particle-hole
(ph) interaction determined as the functional derivative of the
self-consistent mean field with respect to the density. The usual
Tamm-Dancoff approximation (TDA) or RPA in a discrete particle-hole
space amount to solve TDA or RPA equations using the Hamiltonian
$Q_1 H Q_1$.

As said above, giant resonances are known to have a decay width, and we can
distinguish two main mechanisms which give rise to it. The energy of the
vibrational nuclear motion can be transferred out of the system by an
escaping nucleon\footnote{Also a $\gamma$-ray can escape when selection rules
allow it to do. In what follows we neglect this decay channel as it
gives a minor contribution to the total width of the mode.}, or
can be distributed among internal degrees of freedom giving rise to more
complicated configurations than the initial
configuration. The contributions to the
width coming from these two damping effects are usually called respectively
escape ($\Gamma^\uparrow$) and spreading ($\Gamma^\downarrow$) width. We
must also mention another source of broadening of the line width, namely
the Landau spreading which already appears at the level of
discrete RPA or TDA calculations.

In order to account for escape and spreading effects, we build two
other subspaces $P$ and $Q_2$. The space $P$ is made up with
particle-hole configurations where the particle is in an unbound state,
orthogonal to all states $|i\rangle$. To determine these
unbound states, we use the following procedure. At positive energy
$\varepsilon$ we solve, for each partial wave
$c\equiv(l,j)$, the radial scattering equation for $H_0$, projected on the
orthogonal complement of set $|i\rangle$.
We can thus ensure that the
resulting outgoing wave functions
$u^{(+)}_{c,\varepsilon}$ have no overlap
with any of the states $|i\rangle$, i.e., $P$ and $Q_1$ are
orthogonal subspaces (for details, see
\cite{Yos86a,Yos86b,Ada87}). The states
$u^{(+)}_{c,\varepsilon}$ have no resonant behaviour since quasi-bound states
are among the states $|i\rangle$.

The space $Q_2$ is built with a set of ``doorway states'',
the first step in the coupling of the ordered resonance motion
with the compound nuclear states, in which energy is distributed among all
degrees of freedom in a statistical way. We denote
these ``doorway states'' by $|N\rangle$. As
a first approximation we can think of them as formed by 2p-2h
configurations, but we discuss later a more physical choice of them in
terms of states made up with one of the 1p-1h excitations coupled to a
collective vibration.

We decompose the nuclear Green's function $G$ as a sum of terms like
$Q_1GQ_1 + Q_1GP + \cdots$. Using the technique
described in \cite{Yos83} one can
show that by virtue of the equations of motion in the different subspaces
and the properties of projectors, the Green's function $Q_1GQ_1$
obeys an equation
where the effective Hamiltonian, after truncation of higher order couplings,
is
\FL
\begin{eqnarray}
     \ & {\cal H} & (\omega) \equiv Q_1 H Q_1 + W^\uparrow(\omega)
     + W^\downarrow(\omega) \nonumber \\
     = & Q_1 & H Q_1 + Q_1 H P {\textstyle 1 \over \textstyle
     \omega - PHP + i\epsilon} P H Q_1
     + Q_1 H Q_2 {\textstyle 1 \over \textstyle
     \omega - Q_2 H Q_2 + i\epsilon}
     Q_2 H Q_1, \nonumber \\
     \ & \ &
\label{H_eff}\end{eqnarray}
where $\omega$ is the excitation energy.
This energy-dependent, complex Hamitonian allows to work inside the
space $Q_1$. It has complex eigenvalues whose imaginary parts
originate from coupling to unbound and to more complicated
configurations and give rise to escape and spreading widths.

The escape term $W^\uparrow(\omega)$ can be more easily evaluated if
one replaces the complete Hamiltonian $H$ by the one-body part
$H_0$. The neglect of matrix elements of $Q_1V_{ph}P$ should be, in
the present case, rather safe since
discrete and continuum wave functions are essentially
restricted to different radial intervals while $V_{ph}$ interaction has zero
range.
In this approximation, the escape term can be written \cite{Yos86a} as
\begin{equation}
    W^\uparrow(\omega) = \omega - H_0 - K,
\label{Appr_Wup}\end{equation}
where $K$ is the inverse inside subspace $Q_1$ of the Green's function
containing only the mean field Hamiltonian. This Green's function can be
easily computed and inverted as it has only one-body matrix elements whereas
taking into account the two-body interaction would have resulted in a much
harder task. Detailed expressions are given in Appendix A.
The accuracy of the procedure has been checked in \cite{Gia87}, where the
results obtained for the isoscalar monopole strength
distribution in ${}^{40}$Ca, by using the approximate
$W^\uparrow$ of (\ref{Appr_Wup}), were
compared with those of exact continuum-RPA calculations and excellent
agreement between the two calculations was found
(see Fig. 1 of \cite{Gia87}). Yoshida and Adachi made a similar
comparison in \cite{Yos86b}
and they concluded also that a good agreement can be
obtained, provided a sufficiently
large basis is employed (see Figs. 1-3 of \cite{Yos86b}).

The matrix elements on a basis of $Q_1$
can be also determined in a straightforward way for the
spreading term $W^\downarrow(\omega)$. We make the ansatz that
the configurations $|N\rangle$ of $Q_2$ are not interacting\footnote{These
configurations display a rather
high level-density in medium-heavy nuclei, of the order of
10$^2$-10$^3$ levels per MeV in the region of giant resonances if we think
in terms of 2p-2h configurations \cite{Sch84}, so that to calculate their
mutual interaction would be a formidable task.}. This
is reasonable since coupling among doorway states will
correct the coupling between 1p-1h states and doorway states
in higher order of the residual interaction.
If we take a basis of ph
configurations, i.e., a set $|ph\rangle$ (particle levels are labeled by
$p,p^\prime\ldots$, and hole levels by $h,h^\prime\ldots$),
then the generic matrix element turns out to be
\begin{equation}
    W^\downarrow_{ph,p^\prime h^\prime}(\omega) = \sum_N
    {\langle ph | V | N \rangle\langle N | V | p^\prime h^\prime \rangle \over
    \omega - \omega_N},
\label{me_Wdown}\end{equation}
where we have omitted the subscript $ph$ of the interaction, and
$\omega_N$ is the energy of the state $|N\rangle$, in mean field
approximation. Further clarifications about the evaluation of (\ref{me_Wdown})
are given in Sec.\ \ref{sec:chex}.

It is found convenient, in order to solve the effective Hamiltonian
(\ref{H_eff}), to work on the basis of the RPA or TDA states obtained by
diagonalizing its first term. Using this basis, we provide explicit
formulas for the matrix elements of the escape and
spreading terms in Appendix A. In this matrix form, the eigenvalue equation
for the effective Hamiltonian (\ref{H_eff}) is
\begin{equation}
    \left( \begin{array}{cc}{\cal D}+{\cal A}_1(\omega) & {\cal A}_2(\omega) \\
                            {\cal A}_3(\omega) & -{\cal D}+{\cal A}_4(\omega)
                            \end{array} \right)
    \left( \begin{array}{c}  F^{(\nu)} \\ \bar F^{(\nu)} \end{array} \right) =
    (\Omega_\nu - i{\Gamma_\nu\over 2})
    \left( \begin{array}{c}  F^{(\nu)} \\ \bar F^{(\nu)} \end{array} \right),
\label{eig}\end{equation}
where ${\cal D}$ is a diagonal matrix with the RPA eigenvalues, and the
${\cal A}_i$ matrices which are energy dependent contain the escape
and spreading contributions, as explained in Appendix A. We have omitted the
energy dependence of eigenvalues and eigenvectors in (\ref{eig}), for sake
of simplicity in the notation, and
we have collectively denoted
respectively with $F$ and $\bar F$ the amplitudes corresponding
to positive and negative RPA eigenstates.
The solutions of (\ref{eig}) are denoted by $|\nu\rangle$,
whereas the RPA basis vectors are labelled by $|n\rangle$, according to
\begin{equation}
    |\nu\rangle = \sum_n F^{(\nu)}_n |n\rangle.
\label{F_matr}\end{equation}
It is worthwhile at this point to notice that
an approach based on TDA instead of RPA leads to an equation in which
only the upper-left quarter of the matrix that appears in (\ref{eig}) is
present. The matrix which appears in (\ref{eig}) is complex symmetric as can be
seen from its explicit form in Appendix A (see also \cite{Yos86a}). The
transformation which makes it diagonal, that is the matrix of its eigenvectors,
is complex orthogonal, that is
\begin{equation}
    F^T F = FF^T = 1.
\label{FFT}\end{equation}
A useful quantity characterizing a mode excited by a given operator $O$ is
the response function,
\begin{equation}
    R(\omega) = \langle 0 | O^\dagger {\textstyle 1 \over
                \omega -
                {\cal H}(\omega) + i\epsilon} O | 0 \rangle.
\label{resp}\end{equation}
The corresponding
strength function is related to (\ref{resp}) by the well known relation
\begin{equation}
    S(\omega) = -{1\over\pi}Im R(\omega).
\label{str}\end{equation}
In terms of solutions of (\ref{eig}) the strength function is
\begin{equation}
    S(\omega) = -{1\over\pi}Im \sum_\nu \langle 0 | O | \nu \rangle^2
                {\textstyle 1 \over \omega - \Omega_\nu +
                i{\Gamma_\nu\over 2} }\ ,
\label{str_expl}\end{equation}
where the squared matrix element of $O$ appears, instead of its
squared modulus, due to the properties of the eigenvectors $|\nu\rangle$,
which form a biorthogonal basis.
We stress at this point that, although not explicitly written, the eigenvalues
labeled by $\nu$ as well as the wave functions needed to compute the matrix
elements of $O$ depend on $\omega$, as the effective Hamiltonian does.
So, better than the eigenvalue distribution, which can change when
diagonalizing the effective Hamiltonian at different energies,
the strength function really carries information about
the couplings taken into account, and can be directly compared with
experimental results. Note also that, when all the above scheme is carried
out without $Q_2$ space, one should recover continuum RPA results. It is
this kind of consistency check that we alluded to in mentioning
Figs. 1-3 of \cite{Yos86b} and Fig. 1 of \cite{Gia87}.

Another quantity which can be extracted from the model and which is
actually measured in the particle decay experiments is the
branching ratio $B_c$ corresponding to a particular decay channel. An escaping
nucleon with energy $\varepsilon$ leaves a residual $(A-1)$ system in a
hole state such that, by energy conservation $\varepsilon_h =
\varepsilon - \omega$, where $\omega$ is the initial excitation
energy. The cross section $\sigma_c$ for this decay was calculated in
\cite{Yos86a} (see Eq.\  (3.40)), and a discussion is also given in
\cite{Zar85} where a plane wave Born approximation (PWBA) is used to describe
the reaction mechanism. In Appendix B we summarize this
procedure and show how the same PWBA can lead from Eq.\  (3.38)
of \cite{Yos86a} to a more computable expression for the
excitation cross section $\sigma_{exc}$.
Using the results of Appendix B, the branching ratio comes out as
\begin{equation}
    B_c \equiv {\sigma_c \over \sigma_{exc}} =
    \pi i \sum_{\nu,\nu^\prime} { \textstyle
    \gamma_{\nu \nu^\prime, c}
    S_{\nu \nu^\prime} \over \textstyle
    (\omega_{\nu^\prime}-\omega_{\nu})+{i\over 2}(\Gamma_{\nu^\prime}
    +\Gamma_{\nu}) } \left(
    \sum_{\nu,\nu^\prime} (F^* F^T)_{\nu \nu^\prime} S_{\nu \nu^\prime}
    \right)^{-1},
\label{Bc}\end{equation}
where $S_{\nu \nu^\prime}$ is given in (\ref{Sfactor}) and
\begin{equation}
    \gamma_{\nu \nu^\prime,c} = \int d\Omega_k
    \gamma_{\nu,c}({\vec k}) \gamma^*_{\nu^\prime,c}({\vec k}),
\label{ga_int}\end{equation}
with
\begin{equation}
    \gamma_{\nu,c}({\vec k}) = \langle \varphi_c\
    u^{(-)}_{c,\varepsilon}
    ({\vec k}) | H_0 | \nu \rangle.
\label{ga}\end{equation}
In Eq.\  (\ref{ga}),
$\varphi_c$ is the wave function describing the residual $(A-1)$ nucleus in
channel $c$, and $u^{(-)}_{c,\varepsilon}(\vec k)$ is the escaping particle
wave function belonging to $P$ space. It is shown in Appendix B that $B_c$ of
Eq.\  (\ref{Bc}) is a {\em real} quantity.
The strength functions and the branching ratios are the main quantities
we are going to show as results of the present model and which can be
directly confronted with experiment.
Before coming to this, we add some important
remarks relevant to calculations of charge-exchange resonances.

\section{Calculation of charge-exchange resonances}\label{sec:chex}

The model we have just presented was used in its TDA version and
without the inclusion of the spreading term $W^\downarrow$ to
calculate the IAR and GTR in ${}^{208}$Bi \cite{Gia89}. Details
about the HF and TDA procedures as well as the coupling with continuum
configurations can be found in \cite{Gia89}.
On the other hand, the inclusion of the spreading term is an improvement
of the model. Moreover, this term was never included in previous
calculations of the IAR in the form of coupling with collective vibrations, as
it is done here. While coupling to collective vibrations has been used
for the GTR (cf. \cite{Fie82,Bor84}),
the IAR case has only been treated by using for
$|N\rangle$ the 2p-2h configurations to build $W^\downarrow$
(\cite{Ada87,Sch84}). In the present work, we want to use the same model to
construct the $W^\downarrow$ term for IAR and GTR, and therefore
isospin conservation rules must be carefully treated as discussed
at the end of this Sec.

We have consistently used the two interactions SIII and SGII throughout the
whole procedure. The HF mean field for ${}^{208}$Pb is diagonalized on a
basis of 15 shells of the set of harmonic oscillator eigenfunctions with
$\hbar\omega_0$=6.2 MeV (see also \cite{Bla77}). For the construction of
subspace $Q_1$, the set $|i\rangle$ contains all occupied levels and
$\Delta n$ unoccupied proton levels for each value of $(l,j)$.
$\Delta n$ is 6 in the IAR case, and 7 in the GTR case, as in
\cite{Gia89}. The difference with this previous work is the inclusion
in the present case of the Coulomb exchange term in the Slater approximation
inside the HF mean field.

We have then performed a TDA calculation within $Q_1$, as the large neutron
excess of ${}^{208}$Pb should hinder ground-state correlations effects
for proton particle-neutron hole configurations. This gives the basis of
states $|n\rangle$ on which Eq.\  (\ref{eig}) is solved.
We have built the configurations of subspace $Q_2$ by coupling a proton
particle and a neutron hole in the discrete states of the set
$|i\rangle$ with collective vibrations of the ${}^{208}$Pb core.
These doorway states, introduced
above and labeled by $|N\rangle$
play an essential role in the damping of giant resonances.
As pointed out in \cite{Brt83}, the reason is that the choice of 2p-2h
as doorway states amounts to neglect a number of correlations induced
by the residual interaction $V_{ph}$. Many of these correlations are included
if the low-lying collective vibrations are explicitly taken into account as
doorway states.

The collective vibrations of the nucleus ${}^{208}$Pb have been calculated
in self-consistent RPA, with the same SIII or SGII
interaction used throughout the whole work, in a way completely analogous to
\cite{Ber80}, where the coupling with collective vibrations was used to
study the energies and spectroscopic factors of single-particle states.
We refer then for details to
\cite{Ber80}, summarizing here the main points. We have built the complete
spectra of isoscalar modes of $J^\pi$ equal to 2$^+$, 3$^-$ and 4$^+$.
In the configurations of $Q_2$ however, only states with more than 1\% of
the total strength, and with energy less than 20 MeV, were included.
The low-lying $5^-$ state was also included, but its
contribution was found negligible. The collective states included in the
calculations with SIII have been shown in \cite{Ber80}, and those
corresponding to SGII are shown in
Table\ \ref{table:pho} and compared with \cite{Mar86}.
The limits in the accuracy with which this Skyrme-RPA calculation can
reproduce the experimental findings are comparable to the case of
similar calculations performed in the past decades.
At this point we can build the matrix elements of the operator
\begin{equation}
    W^\downarrow(\omega) = \sum_N
    {V | N \rangle\langle N | V \over
    \omega - \omega_N},
\label{op_Wdown}\end{equation}
as in Eq.\  (\ref{me_Wdown}) (these are essential ingredients to build the
matrix elements on TDA basis). $W^\downarrow_{ph,p^\prime h^\prime}$
of (\ref{me_Wdown}) is a sum of terms whose diagrammatic representation is
shown in Fig.\ \ref{fig:dia}. To evaluate these diagrams an expression for the
interaction vertices is needed. This is provided by the particle-vibration
coupling model, by introducing a one-body field which
can be written in the form
\begin{equation}
    V = \sum_{\alpha\beta} \ \sum_{LnM} \langle \alpha |
    \varrho^{(L)}_n (r) v(r) Y_{LM}(\hat r) | \beta \rangle
    \ a^\dagger_\alpha a_\beta.
\label{pvc}\end{equation}
In this equation, we have introduced the radial transition density
$\varrho^{(L)}_n(r)$ of the $|n\rangle$ states of the spectrum of phonons with
angular momentum $L$. This is defined in terms of the RPA states as
\begin{equation}
    \langle n | \psi^\dagger(\vec r) \psi(\vec r) | 0 \rangle \equiv
    \varrho^{(L)}_n (r) Y_{LM} (\hat r),
\label{td}\end{equation}
where $\psi^\dagger(\vec r)$ and $\psi(\vec r)$ are the
creation and annihilation operators of a nucleon at point $\vec r$,
respectively. The form factor $v(r)$ appearing in (\ref{pvc}) is related to the
ph interaction derived from the Skyrme force by
$V_{ph}(\vec r_1,\vec r_2)=v(r_1)\delta(\vec r_1 -\vec r_2)$, as in
\cite{Ber80}. Once all these starting points are given,
the evaluation of the diagrams of Fig.\ \ref{fig:dia} is straightforward
and their detailed expressions are given in Appendix A.

Up to here, the procedure is the one already applied for
the giant monopole resonance in \cite{Col92}. Since we deal here
with charge-exchange modes, it must be noted that the
operator (\ref{op_Wdown}) can mix states with different isospins.
The coupling (\ref{pvc}) is manifestly
a scalar in the total fermion-boson isospin space.
But the intermediate states $|N\rangle$ do not have pure isospin, as
they contain a proton particle and a neutron hole (see the
diagrams of Fig.\ \ref{fig:dia}, where the phonons are isoscalar and
therefore do not
cause any change in the fermion isospin in the intermediate states,
with respect to the initial state). This leads naturally to a coupling of
the IAR, which has isospin quantum numbers $|T,T_z\rangle = |T_0,T_0\rangle$
where $T_0={\textstyle N-Z\over \textstyle 2}$, with states which
in general have a mixture of different $T$ components. The nuclear part of the
Hamiltonian actually forbids this coupling, and we
impose that it is strictly forbidden (this amounts to neglecting Coulomb
effects
in the residual interaction). This can be done by projecting out the
$T_0$ component of the intermediate state $|N\rangle$.
More precisely, we can write in isospin space
\begin{equation}
    |N\rangle = c_{-1} |N;\ T_0-1,T_0-1\rangle +
                c_0    |N;\ T_0,T_0-1\rangle +
                c_{+1} |N;\ T_0+1,T_0-1\rangle,
\label{isoexp}\end{equation}
and we define three projectors $P_{-1}$, $P_0$ and $P_{+1}$ such that
\begin{equation}
    P_i |N\rangle = c_i |N; T_0+i,T_0\rangle, \ -1\leq i \leq 1.
\label{projiso}\end{equation}
The coefficients $c_i$ are simply geometrical factors (Clebsch-Gordan
coefficients),
\begin{eqnarray}
     c_{-1} = & + & (2T_0-1)^{1\over 2}(2T_0+1)^{-{1\over 2}}, \nonumber \\
     c_0    = & - & (T_0+1)^{-{1\over 2}}, \nonumber \\
     c_{+1} = & + & (T_0+1)^{-{1\over 2}}(2T_0+1)^{-{1\over 2}}.
\label{c}\end{eqnarray}
Therefore, in the IAR case the correct spreading operator having the right
isospin structure is
\begin{equation}
    W^\downarrow_{T_0}(\omega) = \sum_N
    {V P_0 | N \rangle\langle N | P_0 V \over
    \omega - \omega_N}.
\label{op_0}\end{equation}
Actually, the matrix elements of this operator are simply proportional to those
of (\ref{op_Wdown}), the multiplicative constant being $|c_0|^2 \equiv
(T_0+1)^{-1}$. In the GTR case, we use the operator
$W^{\downarrow}_{T_0-1}$ defined in a fully analogous way, and thus
proportional
to (\ref{op_Wdown}) by means of the constant $|c_{-1}|^2$.

This procedure allows us to respect isospin symmetry. A similar procedure
was not necessary in the IAR calculations of \cite{Ada87} and \cite{Sch84},
where the intermediate states $|N\rangle$ were chosen as 2p-2h
configurations. The interaction $V$ was the nuclear two-body interaction, and
the approximation used corresponds to the well-known Second RPA (SRPA).
This automatically preserves the
isospin structure, due to the respect of conservation laws in SRPA
\cite{Yan87}, and to the character of the interaction. On the other hand, an
approach similar to ours has been used in \cite{Fie82,Bor84}
for the GTR
without any isospin
factor, i.e., with the spreading operator (\ref{op_Wdown}); as the correct one,
$W^{\downarrow}_{T_0-1}$, differs by a factor $|c_{-1}|^2$ close to 1 in the
case of ${}^{208}$Pb, the numerical results are not markedly affected by this
neglecting of isospin symmetry.

\section{Isobaric analog resonance results}\label{sec:iar}

The calculated strength distributions in the IAR region are shown
in Figs.\ \ref{fig:iarsk3} (interaction SIII)
and\ \ref{fig:iarsg2} (interaction SGII).
Table\ \ref{table:main} includes the
main quantities which can be extracted from these strength distributions, that
is centroid energy, width and percentage of strength of the resonance.
{}From the TDA calculation, a discrete state emerges which, in SIII and
SGII case respectively, lies at 18.59 MeV or 18.54 MeV, and carries 86\% or
87\% of the total strength $(N-Z)$ of the operator $T_- \equiv
\sum_{i=1}^A t_-(i)$. In fact, isospin is explicitly
projected out only in the spreading term, whereas the
TDA calculation is performed in a space with all isospin components included.
Nevertheless, the strength relative to $T_-$ operator selects as most
collective state a single one, which was checked to be almost
pure in isospin by looking at its $T^2$ expectation value. In
Figs.\ \ref{fig:iarsk3} and\ \ref{fig:iarsg2} the dashed line
corresponds to a calculation with $W^\uparrow$ only, and
the full line to the complete
calculation. We remark first that, in contrast with other modes where a certain
amount of fragmentation is present, in this case this strength function is
essentially due to a single state. The effect of $W^\uparrow + W^\downarrow$
is to shift slightly the peak energy and to produce a total width. With SIII,
the peak is at 18.49 MeV and has a width of 152 keV whereas the corresponding
values are 18.64 MeV and 99 keV with SGII. In both cases the state exhausts
97\% of $T_-$ strength. If one performs calculations with $W^\uparrow$ or
$W^\downarrow$ only, one can obtain separately $\Gamma^\uparrow$ and
$\Gamma^\downarrow$. We find that the sum $\Gamma^\uparrow +
\Gamma^\downarrow$ is equal, within a few percent, to the total width
in the complete calculation. It can be seen that this width is rather
sensitive to the interaction used, since the width calculated with SIII is
50\% larger than that of SGII.

The complex collective state, expanded on the TDA basis, shows a squared
overlap of 0.90 (resp. 0.91) with the collective TDA state if one
calculates with SIII (resp. SGII). We must also mention that the spreading term
$W^\downarrow$ has been calculated with an averaging
parameter $i\Delta$ added to the denominator of (\ref{op_0}), for reason of
convenience. This means that one makes a Lorentzian averaging of the
distribution of intermediate
states $|N\rangle$, with $\Delta$ corresponding to half of the Lorentzian
width. The quantity
$\Delta$ has been chosen to be 100 keV, but results are stable
with respect to its variation, as doubling $\Delta$ implies an increase by
only about 20 keV in the total width. Experimental values for IAR excitation
energy and total width are respectively 18.8 MeV and
232 keV \cite{Mel85}.
This states exhausts more or less the whole $T_-$ strength.

Branching ratios corresponding to proton emission leaving the
residual nucleus in a valence
neutron hole state of ${}^{207}$Pb are
shown in Tables\ \ref{table:iarsk3} and\ \ref{table:iarsg2}. In these
Tables, the column labelled ``only $W^\uparrow$'' refers to a
calculation with only the escape included, and the results correspond to
those of \cite{Gia89}. The other three columns under ``$W^\uparrow +
W^\downarrow$'' include the results of the complete calculation. Column (a)
corresponds to the case where the final states are pure Hartree-Fock ones.
These final states can be renormalized by means of energy and spectroscopic
factors, either calculated with the same Skyrme force, and we have these
values in the SIII case in \cite{Ber80} so we can show the
corresponding results for the branching ratios in column (b), or
taken from empirical estimates \cite{Mah91}, and
the branching ratios obtained in this way are shown in column (c). These
calculated results are compared with the experimental values quoted
in Ref. \cite{Aki94}. The agreement
of the total sum of branching ratios with the experimental value in the SIII
case and in
column (b) is quite remarkable, as this is the most consistent fashion
to perform a theoretical calculation. Minor discrepancies are found by looking
in the same column
at individual channels. On the other hand, calculated decay branching
ratios depend sensitively on the Skyrme force employed, even when predicted
peak energy and strength are practically the same.
By examining the IAR wave
function obtained from our calculation, we have noticed that the ph
amplitudes corresponding to a change in the principal quantum number
$n$ are not negligible and this indicates that
a simple picture of the IAR as given by a $T_- |0\rangle$ wave function is
slightly too schematic, at least in the nucleus we have considered.

\section{Gamow-Teller resonance results}\label{sec:gtr}

We follow a similar pattern in presenting the results for the GTR
in ${}^{208}$Bi. Figs.\ \ref{fig:gtrsk3} and\ \ref{fig:gtrsg2} show
the strength functions calculated
respectively with SIII and SGII interactions and
Table\ \ref{table:main} includes
centroids, widths and percentage of strength of these distributions.
With SIII,
at least two main states contribute: besides
the main bump at about 21.50 MeV, a smaller one appears at lower energy. In the
case of SGII, there are two smaller bumps in addition to the main
one at about 23 MeV. One can understand this broadened line shape by looking
at the underlying complex states coming out of the diagonalization of
the effective Hamiltonian (\ref{H_eff}). There are several important complex
states spread over the energy interval 18-25 MeV. This is in contrast with
the IAR case and it can be understood by relating to the larger density of
$T_0-1$ states compared to that of $T_0$ states, in this energy
region of the ${}^{208}$Bi spectrum. The effect of $W^\downarrow$ coupling is
much stronger for GTR and consequently it admixes TDA states more than for
IAR. The percentage of the total strength of the GTR operator
$\beta_- = {1\over 2}\sum_{\mu=0,\pm 1}\sum_{i=1}^A \sigma_\mu(i)\tau_-(i)$
(the so-called Ikeda sum
rule, equal to $3(N-Z)$), is about 61\% in the energy
region 18-24 MeV for SIII, and about 68\% in the interval 19.5-24.5
MeV, for SGII.
This contrasts with the calculation with continuum only, where about the
same fraction of strength is exhausted by
the main narrower peak. In the complete
calculation the remaining strength is found outside the
forementioned interval
($E_<, E_>$) and is rather fragmented.
Experimentally, the strength concentrated in
the GTR peak is about 50\% of the total but this percentage rises
up to about 60-70\% if the whole energy region with some strength around
the main bump is considered \cite{Gaa85}.
All the present theoretical results were calculated with an averaging parameter
$\Delta$ equal to 250 keV. Again, the stability of the results has been
checked by varying $\Delta$.

Besides the peak energy, we can extract from the strength
distribution a mean energy, given by
\begin{equation}
    \langle \omega \rangle =
    {\int_{E_<}^{E_>} d\omega\  \omega S(\omega) \over
     \int_{E_<}^{E_>} d\omega\  S(\omega) },
\label{emean}\end{equation}
and this turns out to be 21.11 MeV (SIII) or 22.43 MeV (SGII).
Both values of $\langle \omega \rangle$ are somewhat larger than
the experimental excitation energy of GTR which is
15.6 MeV in ${}^{208}$Bi, corresponding to 19.2 MeV with respect to
the ${}^{208}$Pb ground
state\footnote{In order to reach this value, we have to add
to the 15.6 MeV the mass difference between the two Bi and Pb isotopes, as
well as the mass difference between neutron and
proton which is missing when the GTR is built on the basis of ph excitation.}.
As for the width, the FWHM is not well defined in this case due to the double
or
triple peak.
If we extract a value for the variance
$\sigma$ from the strength distribution, by defining
\begin{equation}
    \sigma^2 =
    {\int_{E_<}^{E_>} d\omega\  (\omega - \langle
     \omega \rangle)^2 S(\omega) \over
     \int_{E_<}^{E_>} d\omega\  S(\omega) },
\label{variance}\end{equation}
we obtain a full width in Gaussian approximation ($\Gamma=$2.4$\sigma$) of
about 3 MeV for both interactions (slightly larger for SGII),
to be compared with an experimental value of 3.8 MeV
\cite{Aki94}. Of course, a model in which only the simplest
class of ``doorway states''
is considered and not their full hierarchy, cannot but underestimate the
width. The same model seems to predict correctly the
giant monopole width \cite{Col92} which is mainly due to Landau damping.

The overestimation of the peak energy, and the missing part of the damping
effect, is probably the cause of a result for the branching ratios which
correspond to a higher yield than the experimental one, as shown in
Tables\ \ref{table:gtrsk3} and\ \ref{table:gtrsg2}. Here, as the strength is
spread over several MeV, we cannot calculate the branching ratios
at a definite energy, as it was done for IAR, and we rather make an average
over the energy interval ($E_<, E_>$) of the
numerator and denominator of (\ref{Bc}):
\begin{equation}
    B_c ({\rm GTR}) \equiv {\langle \sigma_c \rangle \over \langle \sigma_{exc}
    \rangle}.
\label{bc_ave}\end{equation}
Tables\ \ref{table:gtrsk3} and\ \ref{table:gtrsg2} are displayed similarly to
Tables\ \ref{table:iarsk3} and\ \ref{table:iarsg2} except for the column
(b) of the theoretical findings which we discuss below. First, one
can notice that the discrepancy with experiment is reduced
when going from self-consistent continuum RPA (column ``only W$^\uparrow$'')
to the complete model. The results of SGII are in reasonable agreement
with the data. We must also mention that preliminary indications
from the MSU experiment \cite{Gal_pr} leave open the possibility that
experimental branching ratios might be larger than found in \cite{Aki94}.
On the other hand, the sensitivity of predictions with respect to the choice
of the interaction is still present.
We also show in column (b)
of Tables\ \ref{table:gtrsk3} and\ \ref{table:gtrsg2} what the proton yield
would be,
if the excitation energy corresponded to the experimental one (19.2 MeV).
The branching ratios of column (b)
have indeed been obtained by calculating the escape amplitudes
(\ref{ga}) with the experimental outgoing proton energies $\varepsilon$.
The empirical spectroscopic factors \cite{Mah91} already used in the
case of IAR are also included.
This renormalization
obviously decreases the branching ratios and they are in good agreement
with experiment in the case of SGII force.
The final
difference between theory and experiment, in the case of SIII force, is of
the same order, of even smaller than
the one found in the case of the giant monopole resonance \cite{Col92},
and simply indicates once more the limit one could expect in this kind of
detailed study, due to the present uncertainity in the nuclear dynamics
at low energy.

We can finally compare our results obtained without the spreading terms
with those of Muraviev and Urin \cite{Mur_two}
which are calculated in a continuum-RPA approach, where phenomenological
inputs for the mean field and residual interaction are used.
The large differences between the two sets of results
cannot be explained by invoking the approximation made in the present work
in treating the single-particle continuum, as already discussed
after Eq.\  (\ref{Appr_Wup}). It should be attributed to the sensitivity
of branching ratios to the details of the inputs.
As we have seen here differences in results coming from
the same model but using two different Skyrme interactions, it is not too
surprising that results from phenomenological RPA calculations, based on a
Woods-Saxon potential and a residual interaction of the Landau-Migdal type
can also differ. A similar situation also exists in the giant monopole
resonance case \cite{Bra88}.

\section{Conclusions}\label{sec:conclu}

Within linear response theory, a microscopic model of collective excitations
based on RPA states coupled to doorway states composed of 1p-1h configurations
plus a collective vibration, and to 1p-1h continuum states has been applied
to the study of the properties of two important charge-exchange modes, the
IAR and GTR in ${}^{208}$Bi. The two types of coupling are intended to
describe the essential physical mechanisms leading to the spreading of
the collective mode and its decay by nucleon emission. The model contains
no free parameter but depends on the choice of the effective nucleon-nucleon
interaction.

The physical quantities which can be calculated are the strength
distributions and the cross sections of nucleon decay into various channels.
{}From these quantities centroid energies, widths and particle decay
branching ratios are obtained. The calculated results show some sensitivity
to the choice of the effective interaction.

In the IAR case, both SIII and
SGII interactions predict the correct peak energy, and SIII gives a good
description of the data for the decay branching ratios whereas SGII tends to
underestimate the branching ratios as well as the total width.

For the
GTR, interactions SIII and SGII overestimate the mean energy respectively
by 2 and 3 MeV, while they both predict a total width of about 3 MeV, i.e.,
75\% of the experimental value.
Looking at branching ratios, it appears that SGII comes very close to
experiment
if one corrects for the overestimate of the GTR energy. The
discrepancy with experimental values (in the case of SIII) is of the
same order or even smaller than
that already found in the case of the isoscalar monopole resonance using
the same model. This level of agreement with experiment indicates the
present limits of existing Skyrme effective interactions.

\acknowledgements

One of the authors (G.C.) would like to gratefully acknowledge the support of
ECT*, the continuous interest of R. Leonardi and the useful discussions with
D.M. Brink, in the period November 1993-January 1994, during which a part of
the present work was carried out.

\appendix{Matrix elements of the effective Hamiltonian}

We show in this Appendix how to evaluate matrix elements of the escape and
spreading terms $W^\uparrow$ and $W^\downarrow$ of the effective Hamiltonian
(\ref{H_eff}), on a RPA or TDA basis. We consider explicitly the RPA case, as
the reduction to TDA comes out quite transparently.
First we suppose that the
effective Hamiltonian {\em contains only the RPA and the
escape term}, and we derive the explicit form of Eq.\  (\ref{eig}) in this
case. Then, we show how the spreading term can be taken into account.

We denote a generic element of the RPA basis by $|n\rangle$, its energy being
$\omega_n$. Another state with energy $-\omega_n$ is present in the
basis (see chapter 14 of \cite{Row70}), and we denote it by $|\bar n\rangle$.
The creation operators of these states are respectively
$O_n^\dagger$ and $\bar O_n^\dagger$. The creators ${\cal O}_\nu^\dagger$
of the states
$|\nu\rangle$, resulting from the diagonalization of the effective Hamiltonian
(\ref{H_eff}), can be expressed as a linear combination of the RPA creators,
since these are a complete set in $Q_1$,
\begin{equation}
    {\cal O}_\nu^\dagger = \sum_{\omega_n > 0}
    F^{(\nu)}_n O_n^\dagger - \bar F^{(\nu)}_n \bar O_n^\dagger,
\label{dev_nu_n}\end{equation}
where the amplitudes $F$ and $\bar F$ are in general complex
numbers.

The eigenvalue equation for the effective Hamiltonian is
\begin{equation}
    \left[ {\cal H}, {\cal O}_\nu^\dagger \right] = (\Omega_\nu
    -i{\Gamma_\nu\over 2}) {\cal O}_\nu^\dagger,
\label{eig_eff}\end{equation}
where we have fixed at a definite value the energy at which the effective
Hamiltonian must be evaluated, and omitted to indicate explicitly
this energy dependence. Using the property
\begin{equation}
    \left[ Q_1HQ_1, O_n^\dagger \right] = \omega_n O_n^\dagger,
\label{eig_RPA}\end{equation}
(a minus sign appears in the r.h.s. if we consider $\bar O_n^\dagger$),
we obtain
\FL
\begin{equation}
    \sum_{\omega_n > 0} F^{(\nu)}_n (\omega_n O_n^\dagger
    + [ W^\uparrow, O_n^\dagger ]) - \bar F^{(\nu)}_n (-\omega_n
    \bar O_n^\dagger + [ W^\uparrow, \bar O_n^\dagger ] )
    = (\Omega_\nu-i{\Gamma_\nu\over 2}) \sum_{\omega_n > 0}
    F^{(\nu)}_n O_n^\dagger - F^{(\nu)}_n \bar O_n^\dagger.
\label{eig_dev}\end{equation}
If we take the expectation values of (\ref{eig_dev}), first between
$\langle 0|O_m$ and $|0\rangle$, then between $\langle 0|\bar O_m$ and
$|0\rangle$, we build a set of 2$n_>$ equations ($n_>$ is the number of
positive RPA eigenvalues) whose matrix form is the one of (\ref{eig}).

In (\ref{eig}) each submatrix has dimension $n_>$.
The submatrix ${\cal D}$ is diagonal and contains the contribution of the
term $Q_1HQ_1$ of (\ref{H_eff}),
\begin{equation}
    {\cal D}_{mn} = \delta_{mn} \omega_n.
\label{cal_D}\end{equation}
\widetext
The contributions to the submatrices ${\cal A}_i$ coming from the
escape term are given by
\begin{eqnarray}
     ({\cal A}_1)_{mn}^{{\rm esc}} & = & \langle 0 | \left[ O_m,
        [ W^\uparrow, O_n^\dagger ] \right] | 0 \rangle, \nonumber \\
     ({\cal A}_2)_{mn}^{{\rm esc}} & = & - \langle 0 | \left[ O_m,
        [ W^\uparrow, \bar O_n^\dagger ] \right] | 0 \rangle,
        \nonumber \\
     ({\cal A}_3)_{mn}^{{\rm esc}} & = & \langle 0 | \left[ \bar O_m,
        [ W^\uparrow, O_n^\dagger ] \right] | 0 \rangle, \nonumber \\
     ({\cal A}_4)_{mn}^{{\rm esc}} & = & - \langle 0 | \left[ \bar O_m,
        [ W^\uparrow, \bar O_n^\dagger ] \right] | 0 \rangle.
\label{mel_esc}\end{eqnarray}
To work out explicitly these matrix elements we approximate $W^\uparrow$ with
expression (\ref{Appr_Wup}) (see the discussion in
Sec.\ \ref{sec:gen} about this point).
\narrowtext
The operator $K$ which appears in (\ref{Appr_Wup}) satisfies (as said
in Sec.\ \ref{sec:gen} but see also \cite{Yos86a}) the following equation,
\begin{equation}
    K\cdot Q_1{\textstyle 1 \over \omega - H_0 +i\epsilon}Q_1 = Q_1.
\label{K}\end{equation}
The propagator ${\textstyle 1 \over \omega - H_0 +i\epsilon}$ can
be expanded in its partial
wave components labelled by $(l,j)$ and each component $g_{lj}(r_1, r_2)$
is written by means of the
regular and irregular solutions ($f_{lj}(r)$ and $g_{lj}(r)$) of the
corresponding radial Schr\"odinger equation with energy $\omega$,
\begin{equation}
    g_{lj}(r_1, r_2) = -{2m^* \over\hbar^2} f_{lj}(r_<)g_{lj}(r_>)
    \cdot W^{-1},
\label{schlomo}\end{equation}
where $m^*$ is the nucleon effective mass,
$r_>$ ($r_<$) are the larger (the smaller) between $r_1$ and $r_2$, and
$W$ is the Wronskian of $f_{lj}$ and $g_{lj}$ \cite{Liu76}.
In this form the propagator is more easily inverted to obtain $K$.
This is a ph operator
diagonal in the hole and for a given hole $h$ we label as $K_h$ the
corresponding particle operator.

\widetext
The matrix elements of the escape term are then written in terms of the
ones of $K_h$ and of the RPA amplitudes $X_{ph}^{(n)}$ and $Y_{ph}^{(n)}$ as
\begin{eqnarray}
     ({\cal A}_1)_{mn}^{{\rm esc}} & = - & \sum_{ph} (\varepsilon_p -
        \varepsilon_h) (X^{(m)}_{ph} X^{(n)}_{ph}
        + Y^{(m)}_{ph} Y^{(n)}_{ph}) - \sum_{ph, p^\prime h^\prime}
        \langle p^\prime | K_h | p \rangle \delta_{hh^\prime}
        (X^{(m)}_{p^\prime h^\prime} X^{(n)}_{ph} +
         Y^{(m)}_{p^\prime h^\prime} Y^{(n)}_{ph}), \nonumber \\
     ({\cal A}_2)_{mn}^{{\rm esc}} & = & \sum_{ph, p^\prime h^\prime}
        [(\varepsilon_p - \varepsilon_h) \delta_{pp^\prime}\delta_{hh^\prime}
        + \langle p^\prime | K_h | p \rangle \delta_{hh^\prime}]
        (X^{(m)}_{p^\prime h^\prime} Y^{(n)}_{ph} +
         Y^{(m)}_{p^\prime h^\prime} X^{(n)}_{ph}), \nonumber \\
     ({\cal A}_3)_{mn}^{{\rm esc}} & = & ({\cal A}_2)_{mn}^{{\rm esc}},
     \nonumber \\
     ({\cal A}_4)_{mn}^{{\rm esc}} & = & ({\cal A}_1)_{mn}^{{\rm esc}}.
\label{mel_esc_2}\end{eqnarray}
\narrowtext
We now turn to the calculation of the {\em spreading} term. It is known
(see, e.g., the first chapter of \cite{Spe91}) that diagonalizing an effective
Hamiltonian including this term besides the RPA one, and with a set of
2p-2h configurations in the $Q_2$ space, amounts to
solve
\begin{equation}
    \left( \begin{array}{cc} A + W^\downarrow(\omega) & B \\
                             -B^* & -A^* - W^{\downarrow *}(-\omega)
                            \end{array} \right)
    \left( \begin{array}{c}  X \\ Y \end{array} \right) =
    \omega
    \left( \begin{array}{c}  X \\ Y \end{array} \right),
\label{SRPA}\end{equation}
where $A$ and $B$ are the RPA matrices and $W^\downarrow$ has the matrix
elements defined by (\ref{me_Wdown}). Eq.\  (\ref{SRPA}) is the eigenvalue
problem in the SRPA scheme mentioned in the text. A similar pattern holds
even if a different choice of $Q_2$ is made\footnote{The eigenvalue problem
has the same form but not all the statements about SRPA contained e.g. in
\cite{Yan87} are, strictly speaking, still valid.}, like in the present case,
and therefore, in order to obtain the spreading matrix elements which must be
inserted in (\ref{eig}) one has to transform the spreading part of
Eq.\  (\ref{SRPA}) which is
written on the ph basis of $Q_1$, and write it on the
RPA basis.
\widetext
The result is the spreading contribution to (\ref{eig}), which reads
\begin{eqnarray}
     ({\cal A}_1)_{mn}^{{\rm spr}} & = & \sum_{ph,p^\prime h^\prime}
        W^\downarrow_{ph,p^\prime h^\prime}(\omega) X_{ph}^{(m)}
        X_{p^\prime h^\prime}^{(n)} +
        W^{\downarrow *}_{ph,p^\prime h^\prime}(-\omega) Y_{ph}^{(m)}
        Y_{p^\prime h^\prime}^{(n)}, \nonumber \\
     ({\cal A}_2)_{mn}^{{\rm spr}} & = & \sum_{ph,p^\prime h^\prime}
        W^\downarrow_{ph,p^\prime h^\prime}(\omega) X_{ph}^{(m)}
        Y_{p^\prime h^\prime}^{(n)} +
        W^{\downarrow *}_{ph,p^\prime h^\prime}(-\omega) Y_{ph}^{(m)}
        X_{p^\prime h^\prime}^{(n)}, \nonumber \\
     ({\cal A}_3)_{mn}^{{\rm spr}} & = & \sum_{ph,p^\prime h^\prime}
        W^\downarrow_{ph,p^\prime h^\prime}(\omega) Y_{ph}^{(m)}
        X_{p^\prime h^\prime}^{(n)} +
        W^{\downarrow *}_{ph,p^\prime h^\prime}(-\omega) X_{ph}^{(m)}
        Y_{p^\prime h^\prime}^{(n)}, \nonumber \\
     ({\cal A}_4)_{mn}^{{\rm spr}} & = & \sum_{ph,p^\prime h^\prime}
        W^\downarrow_{ph,p^\prime h^\prime}(\omega) Y_{ph}^{(m)}
        Y_{p^\prime h^\prime}^{(n)} +
        W^{\downarrow *}_{ph,p^\prime h^\prime}(-\omega) X_{ph}^{(m)}
        X_{p^\prime h^\prime}^{(n)}.
\label{mel_spr}\end{eqnarray}
The matrix elements $W^\downarrow_{ph,p^\prime h^\prime}$ can be evaluated
as discussed in Sec.\ \ref{sec:gen}. We give here their final expressions,
\begin{eqnarray}
     W^\downarrow_{ph,p^\prime h^\prime} & = & \sum_{k=1}^4
     W^\downarrow (k) \nonumber \\
     W^\downarrow (1) & = & \delta_{h,h^\prime} \delta_{j_p,j_{p^\prime}}
     \sum_{L,n,p^{\prime\prime}} {1 \over \omega - (\omega_n +
     \varepsilon_{p^{\prime\prime}} - \varepsilon_h) + i \Delta} \cdot
     {|\langle j_p \Vert Y_L \Vert j_{p^{\prime\prime}}
     \rangle|^2 \over {\hat \jmath}^2_p} \nonumber \\
     & \cdot & \int dr_1 u_{p^\prime}(r_1) u_{p^{\prime\prime}}(r_1) v(r_1)
     \varrho^{(L)}_n (r_1)
     \int dr_3 u_p(r_3) u_{p^{\prime\prime}}(r_3) v(r_3)
     \varrho^{(L)}_n (r_3), \nonumber \\
     W^\downarrow (2) & = & \delta_{p,p^\prime} \delta_{j_h,j_{h^\prime}}
     \sum_{L,n,h^{\prime\prime}} {1 \over \omega - (\omega_n -
     \varepsilon_{h^{\prime\prime}} + \varepsilon_p) + i \Delta} \cdot
     {|\langle j_h \Vert Y_L \Vert j_{h^{\prime\prime}}
     \rangle|^2 \over {\hat \jmath}^2_h} \nonumber \\
     & \cdot & \int dr_2 u_{h^\prime}(r_2) u_{h^{\prime\prime}}(r_2) v(r_2)
     \varrho^{(L)}_n (r_2)
     \int dr_4 u_h(r_4) u_{h^{\prime\prime}}(r_4) v(r_4)
     \varrho^{(L)}_n (r_4), \nonumber \\
     W^\downarrow (3) & = & \sum_{L,n} { (-1)^{1+j_{p^\prime}+j_{h^\prime}
     + L + J} \over \omega - (\omega_n +
     \varepsilon_p - \varepsilon_{h^\prime}) + i \Delta}
     \left\{ \begin{array}{ccc}
     j_p & j_h & J \\ j_{h^\prime} & j_{p^\prime} & L \end{array} \right\}
     \langle j_p \Vert Y_L \Vert j_{p^\prime} \rangle
     \langle j_{h^\prime} \Vert Y_L \Vert j_h \rangle \nonumber \\
     & \cdot & \int dr_1 u_p(r_1) u_{p^\prime}(r_1) v(r_1)\varrho^{(L)}_n (r_1)
     \int dr_4 u_h(r_4) u_{h^\prime}(r_4) v(r_4) \varrho^{(L)}_n (r_4),
     \nonumber \\
     W^\downarrow (4) & = & \sum_{L,n} { (-1)^{1+j_{p^\prime}+j_{h^\prime}
     + L + J} \over \omega - (\omega_n +
     \varepsilon_{p^\prime} - \varepsilon_h ) + i \Delta}
     \left\{ \begin{array}{ccc}
     j_p & j_h & J \\ j_{h^\prime} & j_{p^\prime} & L \end{array} \right\}
     \langle j_p \Vert Y_L \Vert j_{p^\prime} \rangle
     \langle j_{h^\prime} \Vert Y_L \Vert j_h \rangle \nonumber \\
     & \cdot & \int dr_2 u_p(r_2) u_{p^\prime}(r_2) v(r_2)\varrho^{(L)}_n (r_2)
     \int dr_3 u_h(r_3) u_{h^\prime}(r_3) v(r_3) \varrho^{(L)}_n (r_3),
\label{mel_spr_det}\end{eqnarray}
where the symbol $\hat \jmath$ denotes $2j+1$.
\narrowtext

\appendix{Decay branching ratios}

In \cite{Yos86a} the cross sections $\sigma_c$ and
$\sigma_{exc}$ (see Sec.\ \ref{sec:gen}) are defined as
\begin{eqnarray}
     \sigma_c & = & {(2\pi)^3 \over k^2}
     \sum_{\nu\nu^\prime} T^*_{\nu 0}T_{\nu^\prime 0} \gamma_{\nu^\prime\nu,c}
     (\omega - \Omega_\nu - i{\Gamma_\nu\over 2})^{-1}
     (\omega - \Omega_\nu + i{\Gamma_{\nu^\prime}\over 2})^{-1},
     \nonumber \\
     \sigma_{exc} & = & -{1\over\pi} {(2\pi)^4\over k^2} Im
     \sum_{\nu\nu^\prime} T^*_{\nu 0} (F^* F^T)_{\nu\nu^\prime}
     (\omega - \Omega_{\nu^\prime} -i{\Gamma_{\nu^\prime}
     \over 2})^{-1} T_{\nu^\prime 0},
\label{sigma_ada}\end{eqnarray}
where all the quantities are defined in the text, apart from
$k$ which is here the wave vector corresponding to the initial state of the
projectile and the elements of the $T$-matrix, in which one state is the
initial state of target (labeled by $0$). These matrix elements are
(see also \cite{Yos86a})
\begin{equation}
    T_{\nu 0} = \langle \chi_f^{(-)} \Phi_\nu | V
    | \Phi_0 \chi_i^{(+)} \rangle,
\label{T_matrix}\end{equation}
where the initial and final wave functions of
the projectile and ejectile are indicated by
$\chi$ and those of the target nucleus by $\Phi$.
Here, $V$ is an effective
projectile-target interaction. We consider for our present purposes
its $\vec\tau\cdot\vec\tau$
($\vec\sigma\cdot\vec\sigma\vec\tau\cdot\vec\tau$) channel for the case of
IAR (GTR) excitation. Either of these operators is labelled by $O$,
so the interaction is $V_O(\vec r_1\ldots\vec r_A; \vec r_p)\cdot O$ where
$\vec r_p$ is the projectile spatial coordinate.
We are in the present work {\em not}
interested in the reaction mechanism, provided it can excite
the target in a state with definite quantum numbers. Therefore,
we can make simplifications about this mechanism and we use standard
PWBA methods to evaluate the element (\ref{T_matrix}). Plane wave functions
are substituted to the initial and final wave functions
of the projectile and ejectile,
and an effective factor $K$
takes care of the reduction of the reaction
probability due to distortion which
would come from a realistic estimate made by using an optical potential.
In a straightforward way (\ref{T_matrix}) becomes
\begin{equation}
    T_{\nu 0} = K\ \tilde V_O(\vec q) \langle \nu
    | e^{{i\over\hbar}\vec q\cdot\vec r}\cdot O | 0 \rangle,
\label{T_PWBA}\end{equation}
where $\vec q$ is the momentum transferred by the projectile and $\tilde
V(\vec q)$ is the Fourier transform of the interaction $V$. The
exponential function can be decomposed as usual,
\begin{equation}
    e^{i\vec q \cdot \vec r} = \sqrt{4\pi} \sum_{\lambda=0}^{\infty}
    (-i)^{\lambda} \sqrt{2\lambda+1} \cdot j_{\lambda} (\vec q \cdot
    \vec r) Y_{\lambda0} ({\hat r}),
\label{exp}\end{equation}
using Bessel functions and spherical harmonics. If we restrict to the
$\lambda$-value under consideration (0 as both IAR and GTR have
no change of angular momentum with respect to the ground state in the
space part of their wave function) and truncate
the expansion of the Bessel function at the lowest order (low momentum
transfer approximation),
we obtain from (\ref{exp}) a constant which can be included in the
definition of $\tilde V$.
Finally we are left with the approximate expression
\begin{equation}
    T_{\nu 0} \sim {\rm constant}\cdot \langle \nu | O | 0 \rangle.
\label{T_appr}\end{equation}
If we use this expression it is easy to show that the branching
ratio, i.e. the ratio between the two cross sections defined in
(\ref{sigma_ada}), is given by (\ref{Bc}) where
\begin{equation}
    S_{\nu\nu^\prime} \equiv \langle \nu | O | 0 \rangle
    \langle \nu^\prime | O | 0 \rangle^*,
\label{Sfactor}\end{equation}
and the other quantities are defined in Sec.\ \ref{sec:gen}.

We conclude this Appendix by proving that the branching ratios defined in
(\ref{Bc}) are real. We show that the complex conjugate,
\begin{equation}
    B_c^* =
    -\pi i \sum_{\nu,\nu^\prime} { \textstyle
    \gamma^*_{\nu \nu^\prime, c}
    S^*_{\nu \nu^\prime} [(\omega_{\nu^\prime}-\omega_{\nu})
    -{i\over 2}(\Gamma_{\nu^\prime}
    +\Gamma_{\nu})]
    \over \textstyle
    (\omega_{\nu^\prime}-\omega_{\nu})^2+
    {i\over 2}(\Gamma_{\nu^\prime}
    +\Gamma_{\nu})^2 } \left(
    \sum_{\nu,\nu^\prime} (F^* F^T)^*_{\nu \nu^\prime} S^*_{\nu \nu^\prime}
    \right)^{-1},
\label{compl_conj}\end{equation}
is equal to $B_c$. Using the fact that the matrix $F$ is complex orthogonal
one obtains
\begin{equation}
    (F^* F^T)^*_{\nu \nu^\prime} = (F^* F^T)_{\nu^\prime \nu}.
\label{FF}\end{equation}
We can insert in (\ref{compl_conj}) this last expression,
together with the analogous
properties of $S$ and $\gamma$, $S^*_{\nu\nu^\prime}=S_{\nu^\prime\nu}$ and
$\gamma^*_{\nu\nu^\prime,c}=\gamma_{\nu^\prime\nu,c}$. Exchanging the
indices $\nu$ and $\nu^\prime$ completes the proof.

\figure{Diagrammatic representation of the four terms whose sum gives the
matrix element W$^\downarrow_{ph,p^\prime h^\prime}$ of (\ref{me_Wdown}).
The analytic expressions are shown in Appendix A,
Eq.\  (\ref{mel_spr_det}). \label{fig:dia}}

\figure{Strength distribution of IAR in ${}^{208}$Bi calculated
with the interaction SIII. The full line refers to the complete calculation,
whereas the dashed line gives the result with only the continuum coupling.
\label{fig:iarsk3}}

\figure{Strength distribution of IAR in ${}^{208}$Bi calculated
with the interaction SGII and displayed as in
Fig.\ \ref{fig:iarsk3}. \label{fig:iarsg2}}

\figure{Strength distribution of GTR in ${}^{208}$Bi calculated
with the interaction SIII. The two lines have the same meaning as in the
case of IAR. \label{fig:gtrsk3}}

\figure{Strength distribution of GTR in ${}^{208}$Bi calculated
with the interaction SGII and displayed as in Fig.\ \ref{fig:gtrsk3}.
\label{fig:gtrsg2}}


\begin{table}
\caption{Vibrational states of ${}^{208}$Pb, resulting from a
coordinate-space RPA calculation employing the SGII force, with energy below
$\sim$ 20 MeV, and exhausted percentage of EWSR larger than 1\%. They are
used to build the ``doorway states'', that is, the configurations
which are more complicated than the 1p-1h configurations and therefore play a
role in the damping process of giant resonances. The experimental values
in this Table are taken from \cite{Mar86}.}
\begin{tabular}{crrrrr}
        &\multicolumn{3}{c}{Theory}
        &\multicolumn{2}{c}{Experiment}\\ \cline{2-6}
J$^\pi$ & Energy & B(E$\lambda$) & Percentage & Energy & B(E$\lambda$) \\
        & [MeV]  & [spu]         & of EWSR    & [MeV]  & [spu]         \\
        &        &               &            &        & (or perc. of EWSR) \\
\tableline
2$^+$   & 4.78   & 3.28          & 7          & 4.085  & 6.2           \\
        & 5.13   & 4.36          & 10         & 4.923  &               \\
        &        &               &            & 5.036  &               \\
        &        &               &            & 5.128  &               \\
        & 11.16  & 15.03         & 75         & 10.6   & (\%EWSR=70)   \\
3$^-$   & 2.64   & 37.68         & 25         & 2.614  & 34.2          \\
        & 13.6   & 0.62          & 1.6        &        &               \\
        & 14     & 0.37          & 1          &        &               \\
        & 16.53  & 0.41          & 1.3        &        &               \\
        & 17.55  & 0.30          & 1          &        &               \\
        & 18.17  & 0.40          & 1.4        &        &               \\
        & 20.68  & 10.13         & 40         & 20.9   & (\%EWSR=36)   \\
4$^+$   & 4.82   & 2.52          & 1.1        & 4.323  & 23            \\
        & 5.71   & 11.21         & 5.8        & 5.690  &               \\
        & 12.05  & 9.61          & 10.5       & 12.    & (\%EWSR=10)   \\
        & 12.44  & 4.52          & 5.1        &        &               \\
        & 13.17  & 2.43          & 2.9        &        &               \\
        & 14.66  & 1.43          & 1.9        &        &               \\
        & 15.14  & 0.80          & 1.1        &        &               \\
5$^-$   & 3.58   & 16.37         & 2.6        & 3.198  & 17.4          \\
\end{tabular}
\label{table:pho}
\end{table}


\begin{table}
\caption{Averaged quantities extracted from the strength distributions
of IAR and GTR. In the case of IAR only one peak appears, as expected, and we
display the energy, the full width and the percentage of strength under
the peak. In the case of GTR we take into account the whole energy region
($E_<, E_>$) in which bumps of the strength function are visible. The
values of $E_<$ and $E_>$, as well as the definitions of the mean energy and
width, are given in the text (Sec.\ \ref{sec:gtr}).}
\begin{tabular}{crrr}
       \multicolumn{4}{c}{Isobaric analog resonance}\\ \cline{1-4}
       & \multicolumn{2}{c}{Theory} & \multicolumn{1}{c}{}\\ \cline{2-3}
       & \multicolumn{1}{c}{SIII}                                &
         \multicolumn{1}{c}{SGII}                                &
         \multicolumn{1}{c}{Experiment \cite{Mel85}}                    \\
\tableline
Mean Energy   & 18.49 MeV   & 18.64 MeV      & 18.8 MeV     \\
Width         & 152 keV     & 99 keV         &  232 keV     \\
Percentage of strength   & 97\%    & 97\%      & $\sim$ 100\%              \\
\tableline
       \multicolumn{4}{c}{Gamow-Teller resonance}\\ \cline{1-4}
       & \multicolumn{2}{c}{Theory} & \multicolumn{1}{c}{}\\ \cline{2-3}
       & \multicolumn{1}{c}{SIII}                                &
         \multicolumn{1}{c}{SGII}                                &
         \multicolumn{1}{c}{Experiment \cite{Aki94}}                    \\
\tableline
Mean Energy   & 21.11 MeV   & 22.43 MeV      & 19.2 MeV     \\
Width         &     3 MeV   &   3.1 MeV         & 3.7 MeV   \\
Percentage of strength   & 61\%    & 68\%      & $\sim$ 60-70\% \cite{Gaa85} \\
\end{tabular}
\label{table:main}
\end{table}


\begin{table}
\caption{Branching ratios for the proton decay of IAR in ${}^{208}$Bi to
neutron valence hole states of ${}^{207}$Pb, obtained by using the SIII force.
In column (a) the final state is a pure HF configuration, whereas in other
columns its energy and wave function are corrected either by means of
a consistent SIII calculation (b) or by using empirical values (c).}
\begin{tabular}{crrrrr}
       & \multicolumn{4}{c}{Theory} & \multicolumn{1}{c}{}\\ \cline{2-5}
Decay  & \multicolumn{1}{c}{only W$^\uparrow$}                   &
         \multicolumn{3}{c}{W$^\uparrow$ + W$^\downarrow$}       &
         \multicolumn{1}{c}{Experiment}                    \\ \cline{2-5}
channel &     & (a)      & (b)         & (c)        & \cite{Aki94} \\
\tableline
p$_{1\over 2}$& 0.472    & 0.346      & 0.253     & 0.237      &0.22$\pm$0.02\\
p$_{3\over 2}$& 0.396    & 0.287      & 0.238     & 0.196      &0.34$\pm$0.04\\
i$_{13\over 2}$& 0.015    & 0.011      & 0.008     & 0.010      & -
\\
f$_{5\over 2}$& 0.117    & 0.086     & 0.065 & 0.061&includ. in p$_{3\over
2}$\\
f$_{7\over 2}$&$<10^{-3}$& $<10^{-3}$ &$<10^{-3}$ &
$<10^{-3}$&0.015$\pm$0.007\\
h$_{9\over 2}$&$<10^{-3}$& $<10^{-3}$ &$<10^{-3}$ & $<10^{-3}$ & -           \\
\hline
$\sum_c B_c$  & 1.       & 0.730      & 0.564     & 0.504
&0.575$\pm$0.07\\
\end{tabular}
\label{table:iarsk3}
\end{table}


\begin{table}
\caption{Same as Table\ \ref{table:iarsk3}, in the case of the SGII force.}
\begin{tabular}{crrrr}
       & \multicolumn{3}{c}{Theory} & \multicolumn{1}{c}{}\\ \cline{2-4}
Decay  & \multicolumn{1}{c}{only W$^\uparrow$}                   &
         \multicolumn{2}{c}{W$^\uparrow$ + W$^\downarrow$}       &
         \multicolumn{1}{c}{Experiment}                    \\ \cline{2-4}
channel &     & (a)      & (c)                & \cite{Aki94} \\
\tableline
p$_{1\over 2}$& 0.448    & 0.171              & 0.137      &0.22$\pm$0.02\\
p$_{3\over 2}$& 0.514    & 0.198              & 0.157      &0.34$\pm$0.04\\
i$_{13\over 2}$& 0.016    & 0.007             & 0.006      & -          \\
f$_{5\over 2}$& 0.016 & 0.008 & -      0.006 &includ. in p$_{3\over 2}$ \\
f$_{7\over 2}$& 0.006    & 0.002              & 0.004 &0.015$\pm$0.007\\
h$_{9\over 2}$&$<10^{-3}$& $<10^{-3}$         & $<10^{-3}$ & -            \\
\hline
$\sum_c B_c$  & 1.       & 0.386              & 0.310   & 0.575$\pm$0.07 \\
\end{tabular}
\label{table:iarsg2}
\end{table}


\begin{table}
\caption{Branching ratios for the proton decay of GTR in ${}^{208}$Bi,
obtained by using the SIII force.
In column (a) the final state is a HF configuration.
In column (b) we present the results of
a calculation where the energy of the final neutron hole state
is corrected in such a way the escaping particle has the same energy as in
the experiment \cite{Aki94}, and spectroscopic factors \cite{Mah91} are used.}
\begin{tabular}{crrrr}
       & \multicolumn{3}{c}{Theory} & \multicolumn{1}{c}{}\\ \cline{2-4}
Decay  & \multicolumn{1}{c}{only W$^\uparrow$}                   &
         \multicolumn{2}{c}{W$^\uparrow$ + W$^\downarrow$}       &
         \multicolumn{1}{c}{Experiment}                    \\ \cline{2-4}
channel &     & (a)      & (b)         & \cite{Aki94}       \\
\tableline
p$_{1\over 2}$& 0.265    & 0.037      & 0.022     & 0.013$\pm$0.002    \\
p$_{3\over 2}$& 0.432    & 0.055      & 0.033     & 0.023$\pm$0.003    \\
i$_{13\over 2}$& 0.009    & 0.001     & 0.001     & 0.002$\pm$0.002    \\
f$_{5\over 2}$& 0.278   & 0.051       & 0.030      &includ. in p$_{3\over 2}$
\\
f$_{7\over 2}$& 0.011    & 0.009      & 0.005     & 0.003$\pm$0.002    \\
h$_{9\over 2}$& 0.005    & 0.001      & 0.001     & -                  \\
\hline
$\sum_c B_c$  & 1.       & 0.154      & 0.092     & 0.041$\pm$0.009      \\
\end{tabular}
\label{table:gtrsk3}
\end{table}


\begin{table}
\caption{Same as Table\ \ref{table:gtrsk3}, in the case of SGII force.}
\begin{tabular}{crrrr}
       & \multicolumn{3}{c}{Theory} & \multicolumn{1}{c}{}\\ \cline{2-4}
Decay  & \multicolumn{1}{c}{only W$^\uparrow$}                   &
         \multicolumn{2}{c}{W$^\uparrow$ + W$^\downarrow$}       &
         \multicolumn{1}{c}{Experiment}                    \\ \cline{2-4}
channel &     & (a)      & (b)         & \cite{Aki94}       \\
\tableline
p$_{1\over 2}$&  0.223   & 0.033      & 0.018     & 0.013$\pm$0.002    \\
p$_{3\over 2}$&  0.418   & 0.035      & 0.019     & 0.023$\pm$0.003    \\
i$_{13\over 2}$& 0.014   & 0.003      & 0.001     & 0.002$\pm$0.002    \\
f$_{5\over 2}$&  0.319   & 0.013      & 0.007     &includ. in p$_{3\over 2}$ \\
f$_{7\over 2}$&  0.016   & 0.010      & 0.003     & 0.003$\pm$0.002    \\
h$_{9\over 2}$&  0.010   & 0.001      & $<10^{-3}$& -                  \\
\hline
$\sum_c B_c$  & 1.       & 0.095      & 0.048     & 0.041$\pm$0.009      \\
\end{tabular}
\label{table:gtrsg2}
\end{table}

\end{document}